\documentclass[twocolumn,prl,showpacs,preprintnumbers,amsmath,amssymb,superscriptaddress,nofootinbib]{revtex4-1}

\usepackage{graphicx}
\usepackage{dcolumn}
\usepackage{bm}
\usepackage{booktabs}
\usepackage{placeins}

\begin{document}


\title{Dawning of the $N=32$ shell closure seen through \\precision mass measurements of neutron-rich titanium isotopes }

\author{E. Leistenschneider}
\email[Corresponding author: ]{erichleist@triumf.ca}
    \affiliation{TRIUMF, 4004 Wesbrook Mall, Vancouver, British Columbia V6T 2A3, Canada}
    \affiliation{Department of Physics \& Astronomy, University of British Columbia, Vancouver, British Columbia V6T 1Z1, Canada}

\author{M.P. Reiter}
    \affiliation{TRIUMF, 4004 Wesbrook Mall, Vancouver, British Columbia V6T 2A3, Canada}
    \affiliation{II. Physikalisches Institut, Justus-Liebig-Universit\"{a}t, 35392 Gie{\ss}en, Germany}  

\author{S. Ayet San Andr\'{e}s}
    \affiliation{II. Physikalisches Institut, Justus-Liebig-Universit\"{a}t, 35392 Gie{\ss}en, Germany}    
\affiliation{GSI Helmholtzzentrum f\"{u}r Schwerionenforschung GmbH, Planckstra{\ss}e 1, 64291 Darmstadt, Germany}

\author{B. Kootte}
    \affiliation{TRIUMF, 4004 Wesbrook Mall, Vancouver, British Columbia V6T 2A3, Canada}
    \affiliation{Department of Physics \& Astronomy, University of Manitoba, Winnipeg, Manitoba R3T 2N2, Canada}

\author{J.D. Holt}
    \affiliation{TRIUMF, 4004 Wesbrook Mall, Vancouver, British Columbia V6T 2A3, Canada}

\author{P. Navr\'{a}til}
    \affiliation{TRIUMF, 4004 Wesbrook Mall, Vancouver, British Columbia V6T 2A3, Canada}

\author{C. Babcock}
    \affiliation{TRIUMF, 4004 Wesbrook Mall, Vancouver, British Columbia V6T 2A3, Canada}

\author{C. Barbieri}
    \affiliation{Department of Physics, University of Surrey, Guildford GU2 7XH, United Kingdom}

\author{B.R. Barquest}
    \affiliation{TRIUMF, 4004 Wesbrook Mall, Vancouver, British Columbia V6T 2A3, Canada}

\author{J. Bergmann}
    \affiliation{II. Physikalisches Institut, Justus-Liebig-Universit\"{a}t, 35392 Gie{\ss}en, Germany}

\author{J. Bollig}
    \affiliation{TRIUMF, 4004 Wesbrook Mall, Vancouver, British Columbia V6T 2A3, Canada}
    \affiliation{Ruprecht-Karls-Universit\"{a}t Heidelberg, D-69117 Heidelberg, Germany}

\author{T. Brunner}
    \affiliation{TRIUMF, 4004 Wesbrook Mall, Vancouver, British Columbia V6T 2A3, Canada}
    \affiliation{Physics Department, McGill University, H3A 2T8 Montr\'{e}al, Qu\'{e}bec, Canada}

 \author{E. Dunling}
    \affiliation{TRIUMF, 4004 Wesbrook Mall, Vancouver, British Columbia V6T 2A3, Canada}
    \affiliation{Department of Physics, University of York, York, YO10 5DD, United Kingdom}

\author{A. Finlay}
	\affiliation{TRIUMF, 4004 Wesbrook Mall, Vancouver, British Columbia V6T 2A3, Canada}
    \affiliation{Department of Physics \& Astronomy, University of British Columbia, Vancouver, British Columbia V6T 1Z1, Canada}

    
\author{H. Geissel}
    \affiliation{II. Physikalisches Institut, Justus-Liebig-Universit\"{a}t, 35392 Gie{\ss}en, Germany}    
\affiliation{GSI Helmholtzzentrum f\"{u}r Schwerionenforschung GmbH, Planckstra{\ss}e 1, 64291 Darmstadt, Germany}

\author{L. Graham}
    \affiliation{TRIUMF, 4004 Wesbrook Mall, Vancouver, British Columbia V6T 2A3, Canada}

\author{F. Greiner}
    \affiliation{II. Physikalisches Institut, Justus-Liebig-Universit\"{a}t, 35392 Gie{\ss}en, Germany}   

\author{H. Hergert}
    \affiliation{National Superconducting Cyclotron Laboratory, Michigan State University, East Lansing, MI 48824,USA}

\author{C. Hornung}
    \affiliation{II. Physikalisches Institut, Justus-Liebig-Universit\"{a}t, 35392 Gie{\ss}en, Germany}

\author{C. Jesch}
    \affiliation{II. Physikalisches Institut, Justus-Liebig-Universit\"{a}t, 35392 Gie{\ss}en, Germany}

\author{R. Klawitter}
    \affiliation{TRIUMF, 4004 Wesbrook Mall, Vancouver, British Columbia V6T 2A3, Canada}
    \affiliation{Max-Planck-Institut f\"{u}r Kernphysik, Heidelberg D-69117, Germany}

\author{Y. Lan}
    \affiliation{TRIUMF, 4004 Wesbrook Mall, Vancouver, British Columbia V6T 2A3, Canada}
    \affiliation{Department of Physics \& Astronomy, University of British Columbia, Vancouver, British Columbia V6T 1Z1, Canada}

\author{D. Lascar}
    \email{Present address: Physics Division, Argonne National Laboratory, Argonne, IL 60439, USA}
    \affiliation{TRIUMF, 4004 Wesbrook Mall, Vancouver, British Columbia V6T 2A3, Canada}

\author{K.G. Leach}
    \affiliation{Department of Physics, Colorado School of Mines, Golden, Colorado, 80401, USA}

\author{W. Lippert}
    \affiliation{II. Physikalisches Institut, Justus-Liebig-Universit\"{a}t, 35392 Gie{\ss}en, Germany}

\author{J.E. McKay}    
    	\affiliation{TRIUMF, 4004 Wesbrook Mall, Vancouver, British Columbia V6T 2A3, Canada}
    \affiliation{Department of Physics and Astronomy, University of Victoria, Victoria, British Columbia V8P 5C2, Canada}

\author{S.F. Paul}
    \affiliation{TRIUMF, 4004 Wesbrook Mall, Vancouver, British Columbia V6T 2A3, Canada}
    \affiliation{Ruprecht-Karls-Universit\"{a}t Heidelberg, D-69117 Heidelberg, Germany}


\author{A. Schwenk}  
    \affiliation{Max-Planck-Institut f\"{u}r Kernphysik, Heidelberg D-69117, Germany}
	\affiliation{Institut f\"ur Kerphysik, Technische Universit\"at Darmstadt, 64289 Darmstadt, Germany}
	\affiliation{ExtreMe Matter Institute EMMI, GSI Helmholtzzentrum f\"ur Schwerionenforschung GmbH, 64291 Darmstadt, Germany}

\author{D. Short}
    \affiliation{TRIUMF, 4004 Wesbrook Mall, Vancouver, British Columbia V6T 2A3, Canada}
    \affiliation{Department of Chemistry, Simon Fraser University, Burnaby, British Columbia V5A 1S6, Canada}

\author{J. Simonis}  
	\affiliation{Institut f\"{u}r Kernphysik and PRISMA Cluster of Excellence, Johannes Gutenberg-Universit\"{a}t, 55099 Mainz, Germany}

\author{V. Som\`{a}}
    \affiliation{IRFU, CEA, Universit\'{e} ́Paris-Saclay, 91191 Gif-sur-Yvette, France}

\author{R. Steinbr\"{u}gge}
    \affiliation{TRIUMF, 4004 Wesbrook Mall, Vancouver, British Columbia V6T 2A3, Canada}

\author{S.R. Stroberg}
    \affiliation{TRIUMF, 4004 Wesbrook Mall, Vancouver, British Columbia V6T 2A3, Canada}
	\affiliation{Reed College, Portland, OR 97202, USA}

\author{R. Thompson}
    \affiliation{Department of Physics and Astronomy, University of Calgary, Calgary, Alberta T2N 1N4, Canada}

\author{M.E. Wieser}
    \affiliation{Department of Physics and Astronomy, University of Calgary, Calgary, Alberta T2N 1N4, Canada}

\author{C. Will}
    \affiliation{II. Physikalisches Institut, Justus-Liebig-Universit\"{a}t, 35392 Gie{\ss}en, Germany}  


\author{M. Yavor}
    \affiliation{Institute for Analytical Instrumentation, Russian Academy of Sciences, 190103 St. Petersburg, Russia}

\author{C. Andreoiu}
    \affiliation{Department of Chemistry, Simon Fraser University, Burnaby, British Columbia V5A 1S6, Canada}


\author{T. Dickel}
    \affiliation{II. Physikalisches Institut, Justus-Liebig-Universit\"{a}t, 35392 Gie{\ss}en, Germany}  
    \affiliation{GSI Helmholtzzentrum f\"{u}r Schwerionenforschung GmbH, Planckstra{\ss}e 1, 64291 Darmstadt, Germany}

\author{I. Dillmann}
	\affiliation{TRIUMF, 4004 Wesbrook Mall, Vancouver, British Columbia V6T 2A3, Canada}
    \affiliation{Department of Physics and Astronomy, University of Victoria, Victoria, British Columbia V8P 5C2, Canada}

\author{G. Gwinner}
    \affiliation{Department of Physics \& Astronomy, University of Manitoba, Winnipeg, Manitoba R3T 2N2, Canada}


\author{W.R. Pla\ss}
    \affiliation{II. Physikalisches Institut, Justus-Liebig-Universit\"{a}t, 35392 Gie{\ss}en, Germany}  
    \affiliation{GSI Helmholtzzentrum f\"{u}r Schwerionenforschung GmbH, Planckstra{\ss}e 1, 64291 Darmstadt, Germany}

\author{C. Scheidenberger}
    \affiliation{II. Physikalisches Institut, Justus-Liebig-Universit\"{a}t, 35392 Gie{\ss}en, Germany}  
    \affiliation{GSI Helmholtzzentrum f\"{u}r Schwerionenforschung GmbH, Planckstra{\ss}e 1, 64291 Darmstadt, Germany}

\author{A.A. Kwiatkowski}
    \affiliation{TRIUMF, 4004 Wesbrook Mall, Vancouver, British Columbia V6T 2A3, Canada}
    \affiliation{Department of Physics and Astronomy, University of Victoria, Victoria, British Columbia V8P 5C2, Canada}  

\author{J. Dilling}
    \affiliation{TRIUMF, 4004 Wesbrook Mall, Vancouver, British Columbia V6T 2A3, Canada}
    \affiliation{Department of Physics \& Astronomy, University of British Columbia, Vancouver, British Columbia V6T 1Z1, Canada}



\date{\today}

\begin{abstract}
A precision mass investigation of the neutron-rich titanium isotopes $^{51-55}$Ti was performed at TRIUMF's Ion Trap for Atomic and Nuclear science (TITAN). The range of the measurements covers the $N=32$ shell closure and the overall uncertainties of the $^{52-55}$Ti mass values were significantly reduced. Our results conclusively establish the existence of weak shell effect at $N=32$, narrowing down the abrupt onset of this shell closure.
Our data were compared with state-of-the-art \textit{ab initio} shell model calculations which, despite very successfully describing where the $N=32$ shell gap is strong, overpredict its strength and extent in titanium and heavier isotones. 
These measurements also represent the first scientific results of TITAN using the newly commissioned Multiple-Reflection Time-of-Flight Mass Spectrometer (MR-TOF-MS), substantiated by independent measurements from TITAN's Penning trap mass spectrometer.

\end{abstract}

\maketitle



Atomic nuclei are highly complex quantum objects made of protons and neutrons. Despite the arduous efforts needed to disentangle specific effects from their many-body nature, the fine understanding of their structures provides key information to our knowledge of fundamental nuclear forces.
One notable quantum behavior of bound nuclear matter is the formation of shell-like structures for each fermion group \cite{Mayer-Jensen}, as electrons do in atoms. 
Unlike for atomic shells, however, nuclear shells are known to vanish or move altogether as the number of protons or neutrons in the system changes \cite{Warner2004}.

Particular attention has been given to the emergence of strong shell effects among nuclides with 32 neutrons, pictured in a shell model framework as a full valence $\nu 2 p_{3/2}$ orbital. Across most of the known nuclear chart, this orbital is energetically close to $\nu 1 f_{5/2}$, which prevents the appearance of shell signatures in energy observables. However, the excitation energies of the lowest $2^+$ states show a relative, but systematic, local increase below proton number $Z=24$ \cite{Xu2015}. This effect, characteristic of shell closures, has been attributed in shell model calculations to the weakening of attractive proton-neutron interactions between the $\nu 1 f_{5/2}$ and $\pi 1 f_{7/2}$ orbitals as the latter empties, making the neutrons in the former orbital less bound \cite{Otsuka2005,Steppenbeck2013}. \textit{Ab initio} calculations are also extending their reach over this sector of the nuclear chart, yet no systematic investigation of the $N=32$ isotones has been produced so far.

Sudden and locally steep drops in the two-neutron separation energies ($S_{2n}$) are also typical indicators of strong shell effects and are accessible through precision mass spectrometry techniques \cite{Blaum2013}.
Mass studies performed at several facilities reveal strong shell effects at $N=32$ in the $_{19}$K \cite{Rosenbusch2015}, $_{20}$Ca \cite{Gallant2012a,Wienholtz2013} and $_{21}$Sc \cite{Xu2015} isotopic chains.
In contrast, the $S_{2n}$ surface is smooth in this region for $_{23}$V and beyond, indicating that the shell has quenched. In fact, spectroscopic data and shell model calculations suggest that the $\nu 1 f_{5/2}$ and $\nu 2 p_{1/2}$ orbitals change their energy order between $_{23}$V and $_{21}$Sc \cite{Liddick2004}.

The picture at the intermediate $_{22}$Ti chain is unclear; presently available data point towards a modest shell effect, but error bars of hundreds of keV, mostly coming from low-resolution or indirect techniques, are not sufficiently small to reveal detailed information, and the data is compatible with the absence of any shell effect within $2 \sigma$. Large deviations have also been observed in the vicinity of Ti after mass measurements were performed using high-resolution techniques \cite{Gallant2012a,Xu2015,Lapierre2012,AME16}, and they enormously impact the current understanding of the local shell evolution. Therefore, precise experimental determination of the mass surface around titanium is necessary to finely understand this transitional behavior.

We present a precision mass survey of neutron-rich titanium isotopes from  mass numbers $A=51$ to 55 performed at TRIUMF's Ion Trap for Atomic and Nuclear science (TITAN) \cite{DILLING2006198}. The measurements probe the $N=32$ shell closure and they are the first systematic investigation of its kind on titanium beyond the $N=28$ shell closure. 
These are also the first scientific results from TITAN using the newly commissioned Multiple-Reflection Time-of-Flight Mass Spectrometer (MR-TOF-MS) \cite{Jesch2015}. The mass determination was also done independently using TITAN's precision Mass measurement PEnning Trap (MPET)  \cite{BRODEUR201220}.


The neutron-rich titanium isotopes were produced through spallation reactions at TRIUMF's Isotope Separator and ACcelerator (ISAC) \cite{Ball2016} facility by impinging a 480 MeV proton beam of 40 $\mu$A onto a low-power tantalum target. The Ti isotopes were selectively ionized using TRIUMF's Laser Ionization Source (TRILIS) \cite{Lassen2009,Takamatsu2015}. The beam was extracted from the target, mass separated at ISAC's high resolution mass separator \cite{Bricault2002} and delivered to the TITAN facility. Besides Ti, the delivered beam typically contained surface-ionized V, Cr, Mn and other lesser produced isobars.

\begin{figure}[ht]
    \begin{center}
        \includegraphics[width=\columnwidth]{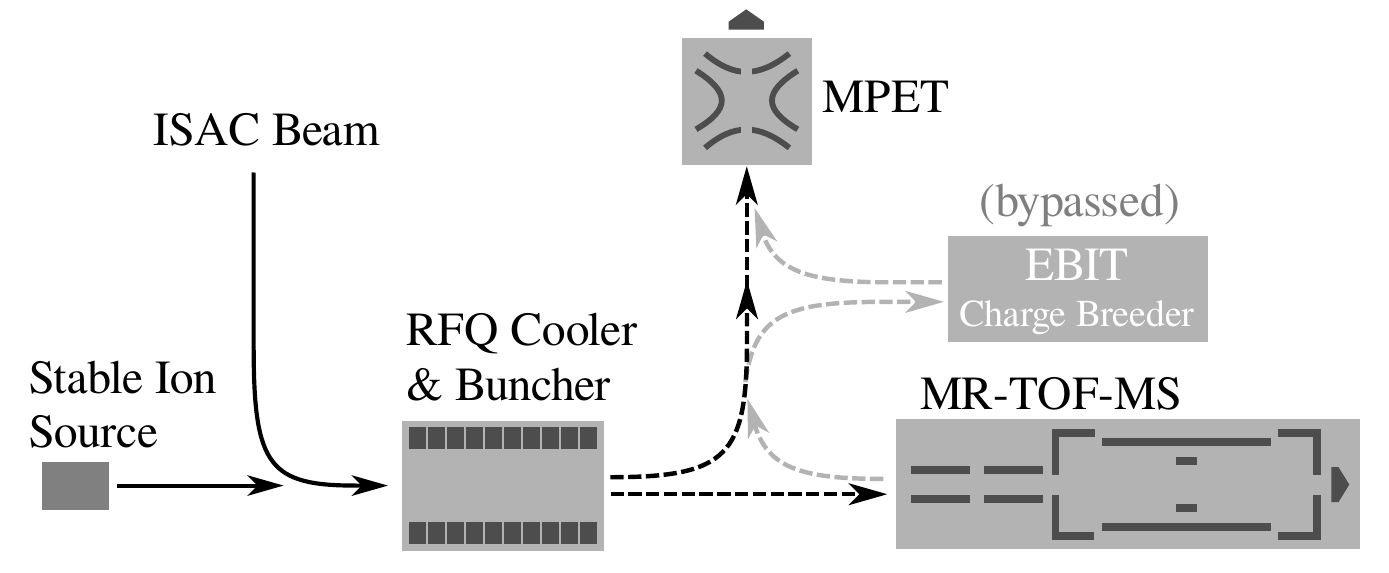}
        \caption{Overview of the TITAN facility highlighting the main components relevant for this experiment. Beam transport of continuous beam is depicted by solid lines and transport of bunched beam is depicted by dashed lines. Transport options not used in this experiment are depicted in light gray.}
        \label{fig:beamline}
        \vspace*{-8mm}
    \end{center}
\end{figure}

At TITAN, the delivered beam was accumulated in a Radio-Frequency Quadrupole cooler and buncher (RFQ) \cite{BRUNNER201232}, which is a preparation trap filled with He gas for cooling. The RFQ can deliver cold bunched beam to the other research stations at TITAN: the MR-TOF-MS, MPET or an Electron Beam Ion Trap charge breeder (EBIT) \cite{Lapierre2010}. This latter unit was bypassed in this experiment. The RFQ can also receive stable beams from TITAN's surface ionization alkali source. An overview of the facility is shown in fig. \ref{fig:beamline}.

For each mass number, the beam delivered by ISAC was cooled in the RFQ and sent in bunches to the MR-TOF-MS for preliminary characterization and mass measurement. Subsequently, in order to validate the mass values of the  MR-TOF-MS calibrants, beam was sent from RFQ to MPET, which is a well established mass spectrometer capable of measuring to higher precision. Mass measurements of both the titanium ion and the chosen MR-TOF-MS calibrant were performed with MPET whenever yields allowed. In this experiment MPET and MR-TOF-MS operated independently, and the details of their measurement techniques are described as follows.


The MR-TOF-MS is a time-of-flight mass spectrometer, in which ions travel a long flight path in a compact setup. Such systems are in operation at ISOLTRAP \cite{Wolf2011}, RIKEN \cite{Schury2009} and FRS at GSI \cite{DICKEL2015} and they are typically able to achieve $10^{-7}$ level of accuracy \cite{AME16}. 
The TITAN device is based on an established concept from the group at the University of Gie{\ss}en \cite{PLASS20084560,PLASS2013134} and is mainly composed of a series of RFQs and RF traps for ion preparation and transport, a time-of-flight mass analyzer and a Micro-Channel Plates (MCP) detector for time-of-flight measurement. 

Beam delivered from the TITAN RFQ was captured in the input RFQ of the MR-TOF-MS and transported to the injection trap system, where it went through another stage of buffer gas cooling. 
The ions were then injected into the mass analyzer, where ion bunches are reflected multiple times between a pair of electrostatic mirrors \cite{YAVOR20151} to provide time-of-flight separation. Inside the mass analyzer, a mass-range-selector \cite{DICKEL2015} was used to deflect any particle outside the desired mass window.   

All MR-TOF-MS mass measurements were done with 512 isochronous turns plus one time-focusing shift turn inside the analyzer for the ions of interest. The time-focusing shift turn \cite{DICKEL2017} was done to adjust the time-focus of the ion bunches to the MCP. The total length of the duty cycle was 20 ms. 
A peak width of about 17 ns was achieved after times-of-flight of about 7.4 ms, corresponding to a mass resolving power of $\approx 220\,000$.

\begin{figure}[ht]
    \begin{center}
        \includegraphics[width=\columnwidth]{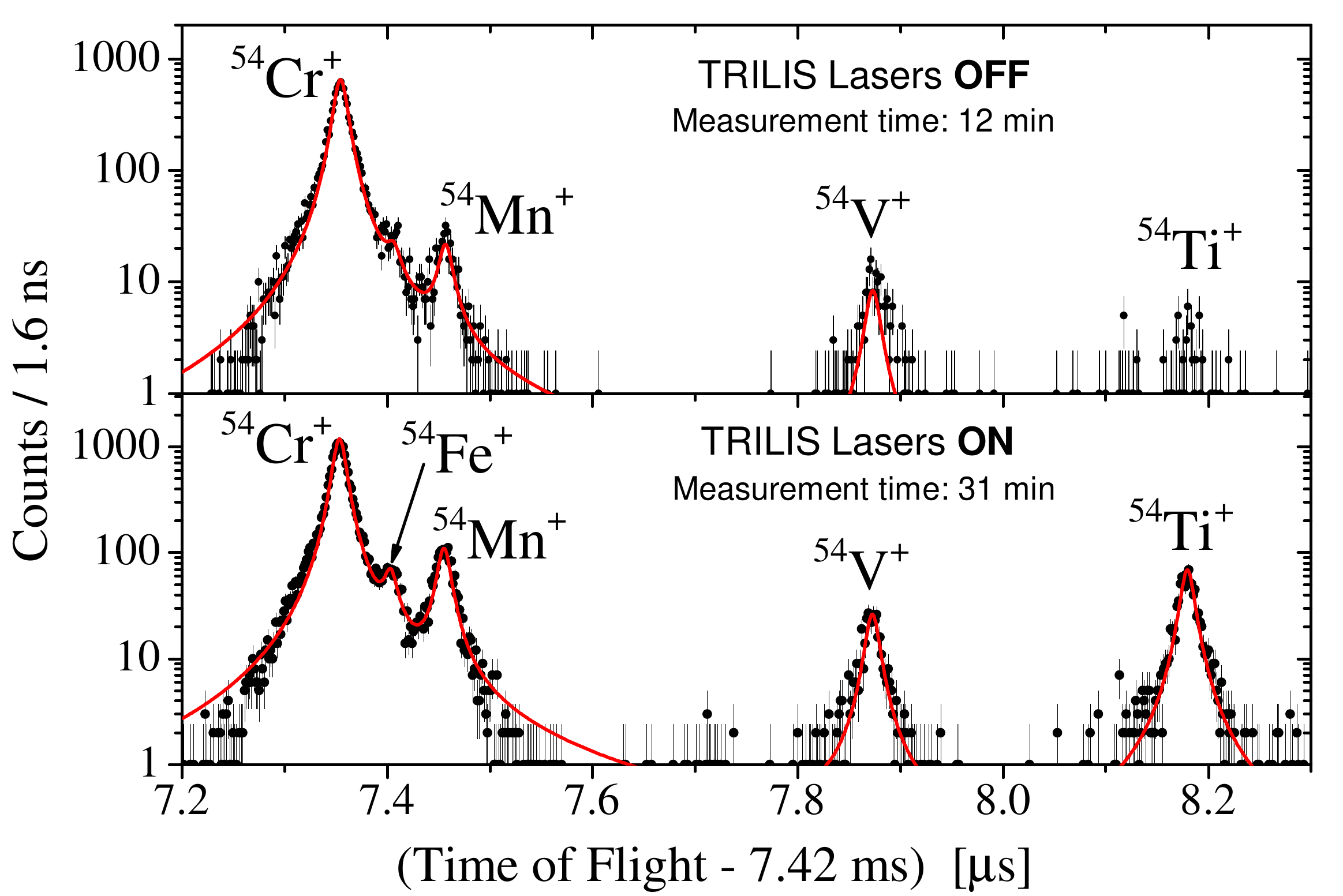}
        \caption{This typical MR-TOF-MS spectrum shows how the identification of titanium peaks was confirmed by turning off the TRILIS lasers. Then, only surface ionized species were delivered to TITAN, causing a sizeable reduction only in Ti yields. In this spectrum, the mass of $^{54}$Ti was determined using the more intense $^{54}$Cr as calibrant. Red curves are fits to the data peaks. }
        \label{fig:mrtofspec}
    \end{center}
\end{figure}

At every mass unit, two measurements were taken: one with the TRILIS lasers switched on and one with the lasers off. This allowed a clear identification of the corresponding Ti peaks in the spectra, as can be seen in figure \ref{fig:mrtofspec}.    
The time-of-flight spectra were corrected for temperature drifts and instabilities in the power supplies by using a time-dependent calibration. 
The peaks were fitted 
and atomic masses $M_a$ were calculated using $M_a = [C (t_{ion} - t_0)^2 + m_e] \, q$, with $m_e$ the rest mass of the electron, $q$ the charge state of the ion, and $t_{ion}$ the fitted time-of-flight centroid of the ion of interest. $C$ is a calibration factor obtained by the mass and time-of-flight of the reference ion, 
while $t_0$ is a small time offset, constant for all measurements and determined from a single turn spectrum using $^{39}$K$^{+}$ and $^{41}$K$^{+}$, prior to the experiment. The uncertainty of the MR-TOF-MS measurements was determined from the statistical uncertainties, the peak forms, and from systematic uncertainties. Systematic contributions were evaluated using both offline \cite{Will2017thesis} and online data to $3\cdot 10^{-7}$, which is dominated by the effects from voltage ringing, 
the uncertainty introduced by the time-dependent calibration, and the presence of overlapping peaks when applicable. A more detailed characterization of MR-TOF-MS' systematic errors will be published in a forthcoming paper. 

Unambiguous identification of titanium was possible in all beams delivered between $A=51$ and 55 and their masses were successfully measured with the MR-TOF-MS. Chromium ions were largely present and were chosen as calibrants for all masses except for $A=51$, in which vanadium was chosen as a more suitable calibrant. 

\vspace*{-5mm}

MPET is a precision Penning trap mass spectrometer dedicated to measuring masses of short-lived unstable isotopes and capable of reaching a $10^{-9}$ level of accuracy \cite{BRODEUR201220}. 
When MPET was used, beam was transported from the RFQ to MPET and injected into the center of the trap, one ion per bunch on average. Ions were prepared for measurement by exciting them onto magnetron motion through the application of a dipolar RF field \cite{Kretzschmar2013}. The major contaminant ions, previously identified through the MR-TOF-MS spectra, were removed through dipolar excitation of the reduced cyclotron motion  \cite{Kretzschmar2013}. The total in-trap ion preparation time was between 60 ms and 70 ms.

The mass measurement is done through the measurement of the ion's cyclotron frequency inside the magnetic field, given by $\nu_c = q \, e \, B/(2\pi M)$, in which $q \, e$ is the charge of the ion, $B$ is the strength of the homogeneous magnetic field and $M$ is the mass of the ion. The procedure employs the well established Time-of-Flight Ion Cyclotron Resonance technique (ToF-ICR) \cite{Konig1995} to measure $\nu_c$. Both standard and Ramsey \cite{George2007} excitation schemes were employed in this experiment, total ToF-ICR excitation times ranged from 100 to 250 ms. 

Every $\nu_c$ measurement of the ions of interest was interleaved by a $\nu_{c,ref}$ measurement of a reference $^{39}$K$^+$ ion, to calibrate the magnetic field and to account for other possible time-dependent variations during the measurement. The atomic mass $M_a$ of the species of interest is calculated from the atomic mass of the reference ion $M_{a,ref}$ and the ratio between their cyclotron frequencies: $R = \nu_{c,ref}/\nu_{c} = (M_a-q \, m_e)/(M_{a,ref}- q \, m_e)$.

We performed mass measurements of $^{51-53}$Ti$^{+}$ and the MR-TOF-MS calibrants $^{51}$V$^{+}$ and $^{52-54}$Cr$^{+}$ using MPET. Yields  were not high enough to perform measurements of $^{54,55}$Ti. 
To characterize any systematic mass-dependent effects, we performed a mass measurement of $^{85}$Rb$^+$, obtained from TITAN's stable ion source. Those were evaluated to be smaller than $1.5 \cdot 10^{-8}$ among the masses of interest, which was included in the error budget. Other known systematic effects \cite{BRODEUR201220,Brodeur2009} 
were evaluated and found to be negligible.



\begin{figure}[ht]
    \begin{center}
        \includegraphics[width=\columnwidth]{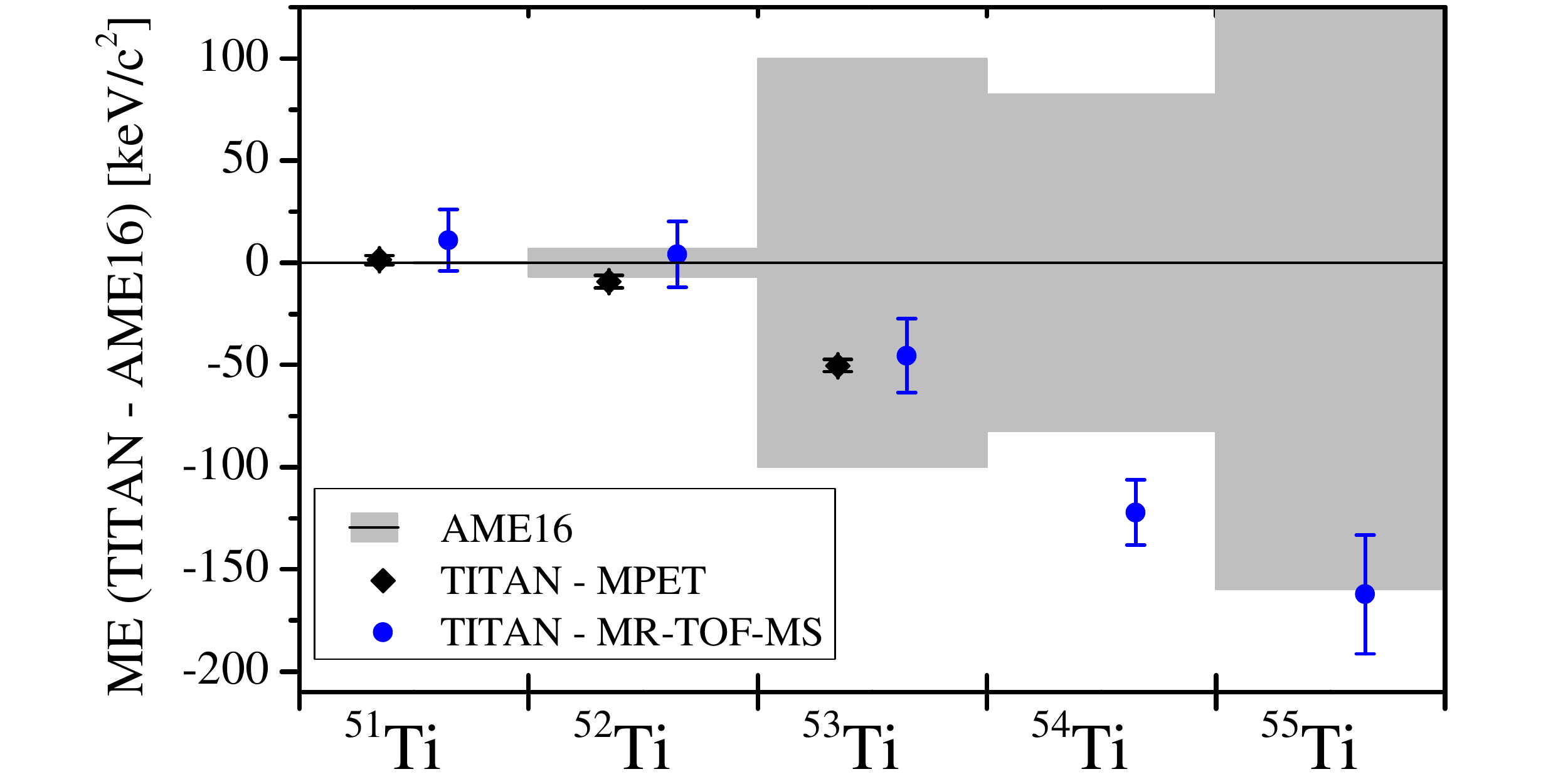}
        \caption{The agreement between MPET and MR-TOF-MS mass measurements can be seen through their mass excesses, plotted here against the AME16 recommended values for comparison. Grey bands represent AME16 uncertainties.   
       }
        \label{fig:massdiff}
        \vspace*{-5mm}
    \end{center}
\end{figure}

All ion species reported were in a singly charged state, therefore atomic mass calculations account for one electron removed. 
Results of all mass measurements performed with MPET and MR-TOF-MS are presented in table \ref{tab:values}, which agree with the Atomic Mass Evaluation of 2016 (AME16) \cite{AME16} recommended values within 1.5$\sigma$ and provide significant reduction of uncertainties. Ti mass excesses are compared against the AME16 values in figure \ref{fig:massdiff} and exhibit a systematic trend towards lower masses for more neutron-rich isotopes. The independent measurements of both spectrometers agree well and were added in quadrature. 

\begin{table}[ht]
  \centering
  \caption{Reported mass measurements performed during this TITAN experimental campaign with the two independent spectrometers: MR-TOF-MS and MPET, and the final TITAN combined values. All MPET mass values are referenced to the mass of $^{39}$K, while references to MR-TOF-MS masses are indicated in the table. Atomic masses are presented as mass excess ($ME$) in keV/$c^2$.  }
    \begin{tabular}{c|c|c|c}
    \toprule
    Species &  $ME_{\text{ MR-TOF-MS}}$  & $ME_\text{ MPET}$ &  $ME_\text{ TITAN}$  \\
    \hline
     $^{51}$V &  \scriptsize{(calibrant)}   &  -52\,203.5 (1.8) & -52\,203.5 (1.8)  \\    
     $^{51}$Ti &   -49\,722 (15)	  &  -49\,731.5 (2.1) & -49\,731.3 (2.1)  \\
    \hline
     $^{52}$Cr &  \scriptsize{(calibrant)}   &  -55\,421.3 (2.0)  & -55\,421.3 (2.0)   \\ 
     $^{52}$Ti &   -49\,466 (16)	&  -49\,479.1 (3.0) & -49\,478.7 (3.0)  \\
    \hline
     $^{53}$Cr &  \scriptsize{(calibrant)}   &  -55288.4 (1.9) & -55\,288.4 (1.9)  \\    
     $^{53}$Ti &   -46\,877 (18)	& -46\,881.4 (2.9) & -46\,881.3 (2.9)  \\
    \hline  
     $^{54}$Cr &   \scriptsize{(calibrant)} &   -56\,929.3 (4.6) & -56\,929.3 (4.6)  \\    
     $^{54}$Ti &    -45\,744 (16)	&  -             & -45\,744 (16)   \\
    \hline     
     $^{55}$Cr &   \scriptsize{(calibrant)} &   -  & -  \\     
     $^{55}$Ti &    -41\,832 (29)	&  -             & -41\,832 (29) \\
    \hline
    \end{tabular}%
  \label{tab:values}%
\end{table}%

\begin{figure*}[ht]
    \begin{center}
        \includegraphics[width=\textwidth]{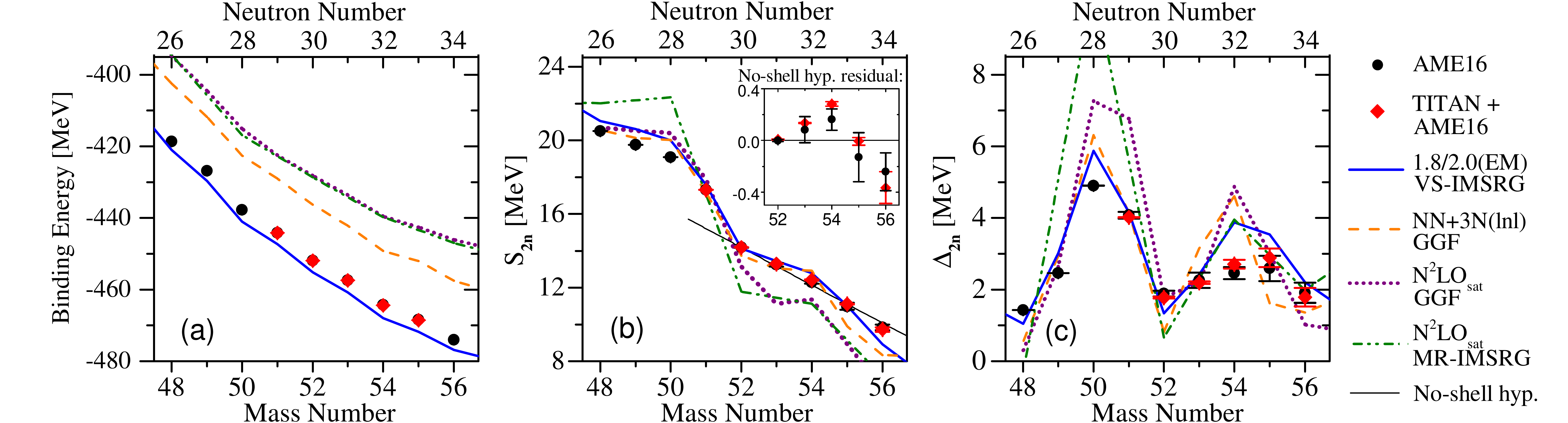}
        \caption{The mass landscape of titanium isotopes is shown from three perspectives: (a) absolute masses (shown in binding energy format), (b) its first ``derivative'' as two-neutron separation energies ($S_{2n}$), and (c) its second ``derivative'' as empirical neutron-shell gaps ($\Delta_{2n}$). Both theoretical \textit{ab initio} calculations (lines) and experimental values (points) are shown. The no-shell hypothesis on $N=32$ is presented in panel \emph{b} as a smooth linear fit to $S_{2n}$ AME16 data between $^{52-56}$Ti, and its residual is shown in the insert, as well as the updated values with TITAN data.}
        \label{fig:shellgapS2n}
        \vspace*{-5mm}
    \end{center}
\end{figure*}

These measurements bring the fine structure of the nuclear mass landscape of the Ti chain to the scale of a few tens of keV.  
In figure \ref{fig:shellgapS2n}(a), titanium binding energies are compared: $BE(N,Z) = M_a(Z,N) - (N\, M_n + Z\, M_p + Z\, m_e )$, where $M_{n,p}$ are the neutron and proton rest masses, respectively. Two ``derivatives'' of the mass landscape are presented in the next two panels: fig. \ref{fig:shellgapS2n}(b) presents the two-neutron separation energies $S_{2n}(N,Z) = M_a(Z,N-2) + 2M_n - M_a(N,Z)$; and panel (c) of same figure shows the empirical neutron-shell gaps $\Delta_{2n}(N,Z) = S_{2n}(N,Z) - S_{2n}(N+2,Z)$, through which shell structures seen in $S_{2n}$ are brought into relief.



The well-known $N=28$ shell closure is easily recognized through the sharp features at $S_{2n}$ and $\Delta_{2n}$ around $^{50}$Ti. Similar but less pronounced characteristics can be seen around $^{54}$Ti, corresponding to the $N=32$ shell. With TITAN data, a no-shell effect hypothesis that assumes a smooth and linear behavior of $S_{2n}$ around $N=32$, once plausible within $2 \sigma$, is completely ruled out by over $50 \sigma$ (see fig. \ref{fig:shellgapS2n}.b and its insert). The measurements presented here conclusively establish the existence of signatures of shell effects at $N=32$ in the Ti chain.

\begin{figure}[ht]
    \begin{center}
        \includegraphics[width=\columnwidth]{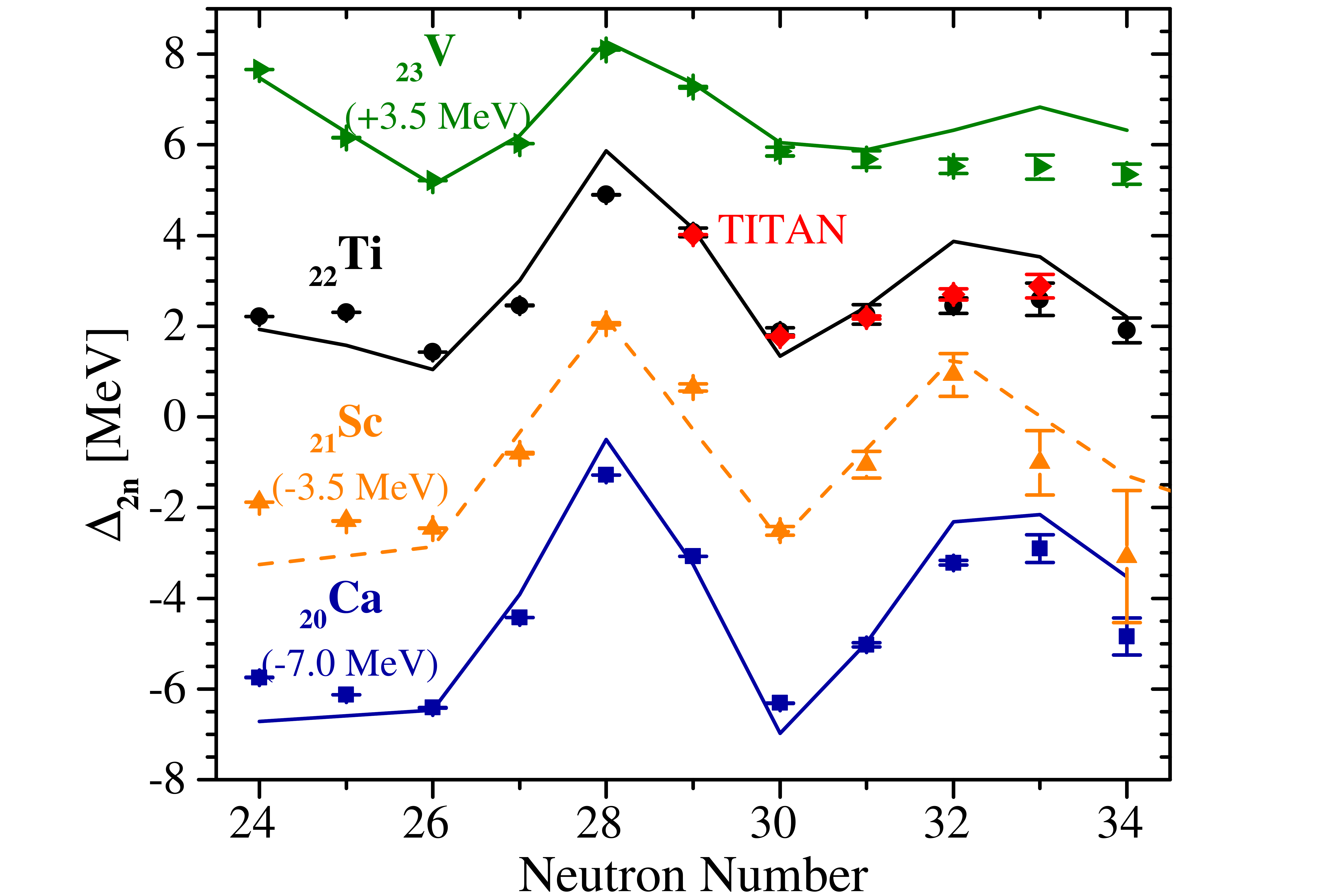}
        \caption{Empirical neutron-shell gaps for titanium and neighboring isotopic chains show the abrupt rise of the $N=32$ shell closure between V and Sc. VS-IMSRG calculations using the 1.8/2.0(EM) interaction (lines) show remarkable overall agreement, but overpredict the extent of the $N=32$ shell closure towards heavier isotones. Data (points) were calculated from AME16 \cite{AME16} values, red data points also include the measurements reported here. Dashed lines in Sc chain are from NN+3N(lnl) GGF calculations. Each isotopic chain was shifted by a multiple of 3.5 MeV for clarity. }
        \label{fig:shellgapall}
        \vspace*{-8mm}
    \end{center}
\end{figure}

The empirical neutron-shell gap at $^{54}$Ti has changed from $2.45(17)$ MeV to $2.70(12)$ MeV, with the mass of $^{56}$Ti now the largest source of uncertainty. In general circumstances, this value alone is no strong indication of a shell closure since the $\Delta_{2n}$ no-shell baseline is approximately 2 MeV in this region. The existence of a special pattern at titanium comes from looking at the $\Delta_{2n}$ systematics with the nearby elements, seen in Fig.~\ref{fig:shellgapall}. It is evident that titanium is at a transition point between V, which shows no signature of a $N=32$ shell closure, and the strong closure seen for Sc and Ca.

With a now clearer picture of the $N=32$ shell evolution, we investigate how well our knowledge of nuclear forces describes the local behaviors. We compared our data to state-of-the-art \textit{ab initio} nuclear structure calculations, shown in Fig.~\ref{fig:shellgapS2n}, based on several nuclear interactions from the recent literature.
In particular, we applied the Multi-Reference In-Medium Similarity Renormalization Group (MR-IMSRG) \cite{Hergert2013a,Hergert2014,Herg16PR}, the Valence-Space (VS-) IMSRG \cite{Tsuk12SM,Bogn14SM,Stro16TNO,Stro17ENO}, and the self-consistent Gorkov-Green’s Function (GGF) \cite{Cipollone2013,Soma2014a,Soma2011,Soma2014b} approaches. 

All calculations were performed with two- (NN) and three-nucleon (3N) interactions \cite{Hebe15ARNPS} based on the chiral effective field theory \cite{Epel09RMP,Mach11PR} with parameters adjusted typically to the lightest systems ($A=2,3,4$) as the only input. In particular, we compare results obtained with the {1.8/2.0(EM)}, the N$^2$LO$_{sat}$ and the NN+3N(lnl) interactions.
The {1.8/2.0(EM)} interaction \cite{Hebeler2011,Simo16unc,Simo17SatFinNuc} combines an SRG-evolved \cite{Bogner2007} next-to-next-to-next-to-leading order (N3LO) chiral NN potential \cite{Entem2003} with a next-to-next-to-leading order (N2LO) non-renormalized chiral 3N force.
The {N$^2$LO$_{sat}$} interaction \cite{Ekstrom2015} has NN and 3N terms fitted simultaneously to properties of $A=2,3,4$ nuclei as well as to selected systems up to $^{24}$O. 
The {NN+3N(lnl)}, applied for the first time in this paper, is a variant of the NN+3N(400) interaction \cite{Roth2012}. It uses both local and non-local 3N regulators (lnl) and refits 3N parameters to $A=2,3,4$ nuclei under a constraint that the contact interactions remain repulsive. The many-body calculations were performed in a harmonic oscillator basis of 14 major shells, with 3N interactions restricted to basis states with $e_{1}+e_{2}+e_{3}\leq e_{\mathrm{3max}} =16$, where $e=2n+l$.

As seen in Fig.~\ref{fig:shellgapS2n}, all approaches were able to predict signatures of shell closures at $N=28$ and $N=32$, although the strength of the neutron shell gap is systematically overpredicted in almost all cases. The calculations with the 1.8/2.0(EM) interaction provide the best description of the Ti data, with masses overbound by only $\approx$ 3.0 MeV, and the neutron shell gaps are closest to the experimentally observed values. The results with the NN+3N(lnl) interaction are also in good agreement with data, though the second order truncation currently employed in GGF calculations results in less total binding energy (typically 10-15 MeV for mid-mass nuclei) compared to more advanced truncation schemes \cite{Duguet2017}. 
The N$^2$LO$_{sat}$ interaction used in the GGF and MR-IMSRG calculations performs well for radii and charge distributions, but here it is found to overpredict the $N=28$ gap compared to 1.8/2.0(EM) and NN+3N(lnl).

Finally, since the VS-IMSRG can access all nuclei in this region, we have employed the 1.8/2.0(EM) interaction to study shell evolution across the known extremes of the $N=32$ shell closure, at Ca (where it is strongest) and V (where it is quenched) isotopic chains, as shown in Fig.~\ref{fig:shellgapall}. First, we see that the calculations provide an excellent description of neutron shell evolution at $N=28$; and, while there is a general overprediction of the neutron shell gap at $N=32$, the trends from $N=28$ to $N=32$ are mostly reproduced. In contrast, calculated shell gaps in titanium steeply rise from $N=30$ to $N=32$ compared to experiment, and even predict modest shell effects in the vanadium chain. This indicates that the $N=32$ closure is predicted to arise too early towards Ca. While the origin of this discrepancy is not completely clear, we note that signatures of shell closures 
are often modestly overestimated by VS-IMSRG \cite{Simo17SatFinNuc}. From direct comparisons with coupled cluster theory \cite{Morr17Sn100}, it is expected that some controlled approximation to include three-body operators in the VS-IMSRG will improve such predictions in magic nuclei and possibly in titanium as well.

In summary, precision mass measurements performed with TITAN's Penning trap and multiple-reflection time-of-flight mass spectrometers on neutron-rich titanium isotopes conclusively establish the existence of weak shell effects at $N=32$, narrowing down the evolution of this shell and its abrupt quenching. 
We also present unprecedented calculations from several \textit{ab initio} theories, including the first ever published results using the NN+3N(lnl) interaction. Overall, all presented theories perform well in this region, but our work reveals deficiencies in the description of the $N=32$ shell if compared to the neighbor $N=28$. Our data provide fine information for the development of the next generation of nuclear forces.
These results also highlight the scientific capabilities of the new TITAN MR-TOF-MS, whose sensitivity enables probing much rarer species with competitive precision.

The authors want to thank the TRILIS group at TRIUMF for Ti beam development, C. Lotze, T. Wasem, R. Wei{\ss} and the staff of the machine shop of the physics institutes of the JLU Gie{\ss}en for excellent technical support.
This work was partially supported by 
Canadian agencies NSERC and CFI, U.S.A. NSF (grants PHY-1419765 and PHY-1614130) and DOE (grant DE-SC0017649), Brazil's CNPq (grant 249121/2013-1), United Kingdom's STFC (grants ST/L005816/1 and ST/L005743/1), German institutions DFG (grants FR 601/3-1 and SFB1245 and through PRISMA Cluster of Excellence), BMBF (grants 05P15RDFN1 and 05P12RGFN8), the Helmholtz Association through NAVI (grant VH-VI-417), HMWK through the LOEWE Center HICforFAIR, and the JLU-GSI partnership.
Computations were performed with resources of the J\"ulich Supercomputing Center (JURECA), GENCI-TGCC (Grant 2017-0507392), MSU’s iCER and UK’s DiRAC Complexity system (grants ST/K000373/1 and ST/K0003259/1). TRIUMF receives federal funding via NRC. 


\bibliography{library}
\end{document}